\documentclass[12pt,reqno]{amsart}

\usepackage{amssymb,amsxtra}
\usepackage{amsfonts}
\usepackage{amsmath} 
\usepackage{graphicx} 
\usepackage{layout}

\newcommand{\erre}{\mathbb{R}} 

\newcommand{\natu}{\mathbb{N}}

\newcommand{\ve}{\varepsilon}

\newcommand{\ma}{\mathfrak{a}}
\newcommand{\mb}{\mathfrak{b}}
\newcommand{\mc}{\mathfrak{c}}

\newcommand{\n}{\noindent} 
\newcommand{\vs}{\vspace{0.5cm}}
\newcommand{\f}{\frac}
\newcommand{\ba}{\begin{eqnarray}} 
\newcommand{\ea}{\end{eqnarray}}
\newcommand{\be}{\begin{equation}}
 \newcommand{\ee}{\end{equation}}
\newcommand{\bdm}{\begin{displaymath}}
\newcommand{\edm}{\end{displaymath}} 
\newcommand{\brr}{\begin{array}}
\newcommand{\err}{\end{array}}

\newcommand{\bml}{\begin{gather}} \newcommand{\eml}{\end{gather}}

\newcommand{\spaz}{\vspace{.5cm} \noindent}

\numberwithin{equation}{section}

\begin{document}


\vs

\title[ ]{A time-dependent perturbative analysis for a quantum particle in a cloud chamber
\\ \vspace*{0.6cm}}

\author{Gianfausto dell'Antonio}
\address{ G. Dell'Antonio: \newline
Dipartimento di Matematica, Universit\'a di Roma "La Sapienza",  
P.le A. Moro, 2 - 00185 Roma (Italy)  and S.I.S.S.A., via Beirut, 2-4  -  34151 Trieste (Italy) \newline
{\em E-mail: gianfa@sissa.it}}



\author{Rodolfo Figari} 
\address{ R. Figari: \newline Dipartimento di Scienze Fisiche,
Universit\'a di Napoli and Sezione I.N.F.N. Napoli, Via Cinthia, 45 - 80126 Napoli (Italy) \newline {\em E-mail: figari@na.infn.it}}


\author{Alessandro Teta} 
\address{ A. Teta: \newline Dipartimento di Matematica Pura ed
Applicata, Universit\`a di L'Aquila,  Via Vetoio, loc. Coppito  -  67010 L'Aquila (Italy) \newline {\em E-mail: teta@univaq.it}}


{\maketitle}

\vs

\begin{abstract}
We consider a simple model of a cloud chamber consisting of a test particle (the $\alpha$-particle) interacting with two other particles (the atoms of the vapour)  subject to attractive potentials centered in $a_1, a_2 \in \mathbb{R}^3$.
At time zero the $\alpha$-particle is described by an outgoing spherical wave centered in the origin and the atoms are in their ground state.
We show that, under suitable assumptions on the physical parameters of the system and up to second order in perturbation theory, the probability that both atoms are ionized is negligible unless $a_2$ lies on the line joining the origin with $a_1$.
The work  is a fully time-dependent version of the original analysis proposed by Mott in 1929.

\end{abstract}

\vs

\section{Introduction}

\vs
\n
The classical limit of quantum mechanics is a widely studied subject in mathematical physics and many detailed results on the asymptotic regime $\hbar \rightarrow 0$ are  available (see e.g. \cite{r} and references therein).  Nevertheless, in most cases, the limit $\hbar \rightarrow 0$ for the time-dependent Schr\"{o}dinger equation relative to a given quantum system is studied only for suitable chosen, "almost classical" states, namely WKB or coherent states. Roughly speaking, these results guarantee that if one choses an almost classical initial state then for $\hbar \rightarrow 0$ its propagation remains close to a classical propagation   at any (not too long) later time. The problem arises when one considers a situation in which the quantum system at initial time is in a genuine quantum state, e.g. a superposition state, and  nevertheless the system exhibits a classical behavior. Examples of this situation are  the localization effect in chiral molecules or the suppression of the interference fringes for a heavy particle in a two-slit experiment. It is clear that  the emergence of such classical behavior cannot be understood if one insists to consider the limit $\hbar \rightarrow 0$ for the isolated quantum system. It is worth mentioning  that the problem has some relevance from the  conceptual point of view.  In fact it was already raised 
in the earliest  debate on the foundation of Quantum Mechanics (see e.g. \cite{be}).  The accepted explanation in these cases is based on the consideration of the interaction of the quantum system with an environment, and in particular on the decoherence effect produced by the environment. It is important to stress that the decoherence effect must be proved in each situation, starting from specific models of system + environment and introducing suitable assumptions on the parameters of the model. Many results in this direction have been obtained in the physical literature (see e.g. the reviews \cite{gjkksz}, \cite{horn}  and references therein) but only few mathematical results are available.

\n
Here we want to focus on an old problem of a different kind, raised by Mott (\cite{m}) in $1929$, concerning the explanation of the straight tracks left by an $\alpha$-particle in a cloud chamber. According to quantum mechanics (\cite{ga}, \cite{cg}) the $\alpha$-particle, isotropically emitted by a radioactive source, is initially described by a spherical wave function and then interacts with the atoms of the vapour, whose positions are randomly distributed in the chamber. The observed tracks are the macroscopic manifestation of the ionization of the atoms induced by the $\alpha$-particle and "it is a little difficult to picture how it is that an outgoing spherical wave can produce a straight track; we think intuitively that  it should ionise atoms at random throughout space" (\cite{m}). The explanation proposed by Mott was based on a simple model describing the $\alpha$-particle in interaction with only two atoms. Exploiting time-independent perturbation arguments, he concluded that the probability that both atoms are ionized is negligible unless the two atoms and the center of the spherical wave lie on the same line. In such rather indirect sense, this would explain why we see straight tracks in the experiments. We will remark on this aspect in section 3.  Notice that so far we have spoken of tracks {\em due} to the decay of an $\alpha$-particle.   In order to be able to speak of a track {\em of} an $\alpha$-particle one must use the reduction postulate which states that the initial spherical wave becomes localized at the place where the ionization takes place. 

\n
We mention that the problem is also discussed in \cite{h} and later in \cite{b}, and some further elaborations on the subject can be found in \cite{p}, \cite{br}, \cite{ccf}, \cite{cl}, \cite{ha}, \cite{st}.

\n
In this paper we reconsider the three-particle model of a cloud chamber. Under suitable assumptions on the parameters of the model, which will be specified later, we give a proof of Mott's result through a fully  time-dependent analysis and up to second order in perturbation theory. The method of the proof is rather elementary and it basically relies on stationary and non stationary phase arguments for the estimate of the oscillatory integrals appearing in the perturbative expansion.  The work extends to the three-dimensional case the result obtained in \cite{dft} for the simpler one-dimensional case, where the spherical wave reduces to  the coherent superposition of two wave packets with opposite mean average momentum.

\vs

\vs
\section{description of the model
}

\vs

\n
In this section we describe the model, i.e. the Hamiltonian, the initial state, the assumptions on the physical parameters, and introduce some notation. 

\n
Let us introduce the Hamiltonian. 
We consider a three-particle non relativistic, spinless  quantum system in dimension three, made of  a  particle   with mass $M$ (the $\alpha$-particle) and two other particles  with   mass $m$ 
 which play the role of model-atoms. More precisely we describe such atoms
as particles  subject to an attractive point interaction placed at fixed positions $a_1\, ,\, a_2 \in \erre^3$, with $|a_1| < |a_2|$ (as we shall see later, the result is still valid if the point interaction is replaced by the Coulomb potential). Moreover  we assume that the interaction between the $\alpha$-particle and each atom   is given by a smooth two-body potential $V$. 
We denote by $R$ the position coordinate of the  $\alpha$-particle and by $r_1$,$r_2$ the position coordinates of the two atoms. 
The Hamiltonian of the system  in $L^2(\erre^9)$ is formally written as
\ba
&&H = H_0 + \lambda H_1 \label{H}\\
&&\nonumber\\
&&H_0= K_0   + K_1 + K_2   \label{H_0}\\
&&\nonumber\\
&&H_1 = V(\gamma^{-1}(R-r_1)) + V(\gamma^{-1}(R-r_2)) \label{H_1}
\ea
where $K_0$ denotes the free Hamiltonian for the $\alpha$-particle 
\ba\label{k0}
&&K_0 = -\f{\hbar^2}{2M} \Delta_R\, ,
\ea

\n
$\lambda >0$ is a coupling constant and $K_j$, $j=1,2$, is the Schr\"{o}dinger operator in $L^2(\erre^3)$ with an attractive point interaction of strength $-(4 \pi \gamma)^{-1}$, $\gamma >0$,  placed at $a_j$. 
The operator $K_j$ is by definition a non trivial selfadjoint extension of the free Hamiltonian restricted on $C_0^{\infty}(\erre^3 \setminus \{a_j\} )$   (for further details see e.g. \cite{al}). The spectrum is 

\be\label{spe}
\sigma_{p} (K_j) = \left\{E_0  \right\}, \;\;E_0 = - \f{ \hbar^2}{2 m  \gamma^2},\;\;\;\; \;\;\;\;\;\; \sigma_c (K_j)=\sigma_{ac} (K_j) = [0, \infty) 
\ee

\n
and the proper and generalized eigenfunctions are respectively given by

\ba\label{autp}
&&\zeta_j (r) = \f{1}{\gamma^{3/2}} \, \zeta^0 (\gamma^{-1} (r-a_j))\,, \;\;\;\;\;\;\;\;\;\;\;\;\;\;\; \zeta^0 (x)=   \f{1}{\sqrt{2 \pi }}\, \, \f{e^{-   |x|}}{|x|}  \\
&&\phi_j (r,k)= e^{i k \cdot a_j} \phi^0 (\gamma^{-1}(r-a_j), \gamma k)\,, \;\;\;\;\; \phi^0 (x,y)\!= \!\f{1}{(2 \pi)^{3/2}}\! \left( \!   e^{i y \cdot x}   -  \f{1}{1\! -  i  |y| } \f{e^{- i |y||x|}}{|x|} \!\right)  
\label{autg}
\ea

\n

\n

\n
The parameter $\gamma$ has the physical meaning of a scattering length and it characterizes  the effective range of the point interaction.  From (\ref{autp}) it is also clear that $\gamma$  is a measure of the linear dimension of the ground state, i.e. of the atoms.



\n
The unperturbed Hamiltonian $H_0$ is obviously selfadjoint and bounded from below in $L^2(\erre^9)$ and moreover the smoothness assumption on the interaction potential $V$  (see theorems 1, 2) guarantees that  the perturbed Hamiltonian $H$ is also selfadjoint and bounded from below on the same domain of $H_0$. In particular this implies that the evolution problem associated with the Hamiltonian $H$ is well posed. 

\n
We choose the initial state in the product form

\be\label{in}
\Psi_0 (R, r_1, r_2) = \psi (R) \zeta_1 (r_1) \zeta_2 (r_2)
\ee
where $\zeta_j$ is defined in (\ref{autp}) and   $\psi(R)$ is a spherical wave  defined as follows. Let us consider a gaussian wave packet localized in space around the origin, with standard deviation $\gamma$ and  mean momentum $P_0>0$ along the direction $\hat{u} \in S^2$. Integrating over the unit sphere $S^2$, one obtains

\be
\psi (R) = \f{\mathcal{N}}{\ve \, \gamma^{3/2}} \,f(\gamma^{-1}R)  \int_{S^2} \!\! \! d\hat{u} \; e^{\f{i}{\ve} \hat{u}\cdot \f{R}{\gamma}} \;, \;\;\;\;\;\;\;\; f(x)= e^{-\f{|x|^2}{2}}
\label{sfe}
\ee

\n
where $\mathcal{N}$ is a normalization factor  and $\ve >0$ is the dimensionless parameter

\be\label{eps}
\ve \equiv \f{\hbar}{P_0 \gamma}
\ee

\n
By an elementary integration we  get

\be\label{sfe2}
\psi(R)= \f{4 \pi  \mathcal{N}}{\gamma^{1/2}} \, \f{e^{-\f{R^2}{2\gamma^2}}}{|R|}  \sin (\ve^{-1} \gamma^{-1}|R|) , \;\;\;\;\;\;\;\;\;\;\mathcal{N}= \f{1}{4 \pi^{7/4}   \left(\!1-e^{-\f{1}{\ve^2}} \!\right)^{\!1/2}}
\ee

\n
We remark that the characteristic length $\gamma$ appears in the definition of the Hamiltonian as well as in the initial state in such a way that  the range of the interaction between the $\alpha$-particle and the atoms, the linear dimension of the atoms and the localization in space of the spherical wave  
are all of order $\gamma$. As it will become clear further on, this is a crucial ingredient for the proof of our result.

\n
We also notice that  the initial state (\ref{in}) corresponds to the situation considered by Mott, i.e. the $\alpha$-particle emitted as  a outgoing spherical wave and the atoms in their ground state.

\n
Let us describe the  hypotheses  on the physical parameters of the model. We assume
\be\label{ve}
\ve  \ll 1
\ee

\be\label{siga}
\f{\gamma}{|a_j|} =O(\ve), \;\;\;\;\;\;\;\; j=1,2
\ee

\be\label{mM}
\f{m}{M} = O(\ve)
\ee

\n
Condition (\ref{ve}) means that 
the wavelength $\hbar P_0^{-1}$ associated to the initial state of the $\alpha$-particle is much smaller than the  linear dimension of the atoms and the range of the interaction,  which means that we are in a semi-classical regime for the $\alpha$-particle. 
In (\ref{siga}) we assume that $|a_1|$, $|a_2|$ are macroscopic distances with respect to the characteristic length $\gamma$ 
and in (\ref{mM}) we require that the mass ratio is small.  
Finally we tacitly  assume  
\be\label{la0}
 \lambda_0 \equiv \f{\lambda}{M v_0^2}= O(\ve)
\ee
where $v_0= P_0 M^{-1}$. Condition (\ref{la0}) is necessary in order to make reasonable the application of our perturbative techniques, even if it is not strictly required for the proof of our results.  
The above assumptions (\ref{ve}), (\ref{siga}), (\ref{mM}) have some relevant physical implications. In particular from (\ref{ve}), (\ref{mM}) one sees that the binding energy  of the atoms is  small compared to the kinetic energy  of the $\alpha$-particle

\ba\label{mE}
&&\f{2|E_0|}{Mv_0^2} = 
 \f{M}{m} \left(\! \f{\hbar}{P_0 \gamma} \!\right)^{\!\!2}=
O(\ve)
\ea

\n
Furthermore the assumptions (\ref{ve}) and (\ref{siga}) imply two  relations among the characteristic times of the system which will be relevant in the what follows. In particular we define the flight times to the atoms of the $\alpha$-particle
\ba\label{ft}
&&\tau_j = \f{|a_j|}{v_0}, \;\;\;\;\;\;\; j=1,2
\ea
the characteristic "period" of the atoms
\ba\label{ta}
&&T_a = 2 \pi \f{ \hbar}{|E_0|} = 4 \pi\f{m \gamma^2}{\hbar}
\ea
and the transit time of the $\alpha$-particle in the region where the atom are localized

\ba\label{tt}
&&T_t = \f{\gamma}{v_0}
\ea

\n
Then one has

\ba\label{tttj}
&&
\f{T_t}{\tau_j} = \f{\gamma}{v_0} \f{v_0}{|a_j|}=  O(\ve), \;\;\;\;\;\;\;j=1,2 \\
&& \f{T_t}{T_a} = 
\f{1}{4 \pi} \f{M}{m} \f{\hbar}{P_0 \gamma}
=O(1)
\label{ttta}
\ea

\n
i.e.  the transit time $T_t$ is small with respect to the flight times $\tau_j$ but it is comparable with the characteristic period of the atoms $T_a$. This means that the $\alpha$-particle can "see" the internal structure of the atoms.

\n
Let us introduce some  notation  to streamline the presentation.

\ba\label{ah}
&&\hat{a}_j = \f{a_j}{|a_j|}, \;\;\;\;\; j=1,2
\\
&&\label{ome}
\omega (y) = \f{1}{2} (1 + y^2), \;\;\;\;\;y \in \erre^3 \\
\label{frak}
&&\ma= \f{\hbar t}{M\gamma^2}, \;\;\;\;
\mathfrak{b}_j = \f{\hbar \tau_j}{M \gamma^2}, \;\;\;\;  \mathfrak{c}_j = \f{\hbar^2 t}{P_0 \gamma \, m \gamma^2}\, \omega(y_j), \;\;\;\;\;j=1,2 \\
&&h(\xi,y)= \f{1}{(2 \pi)^{3/2}} \int \!\! dx \, e^{-i \xi \cdot x} \, \overline{\phi^0} (x,y) \, \zeta^0(x) , \;\;\;\;\; \xi, y \in \erre^3 \label{h}\\
\label{g}
&&g(\xi, y)= \widetilde{V}(\xi)      h(\xi,y)
\ea

\n
where $\tilde{F}$ denotes the Fourier transform of $F$.  Moreover we denote by $\|\cdot \|_{W^{p,n}_{m}}$, $p\geq1$, $n,m \in \natu$, the standard weighted Sobolev norm in $\erre^3$, and by $\mathcal{S}(\erre^3)$ the Schwartz space of real function in $\erre^3$. Finally $\mathcal{C}_k$ denotes a positive numerical constant, depending on $k \in \natu$ and, possibly, on the dimensionless parameters (\ref{frak}).

\n
It is important to notice that, for any fixed $t>\tau_2$ and $y_j \in \erre^3$, the quantities $\ma, \mb_j, \mc_j$ are all of order one. In fact it is sufficient to notice that
\ba\label{ma}
&&\ma = \f{\hbar}{P_0 \gamma} \f{|a_2|}{\gamma} \f{t}{\tau_2},  \;\;\;\; \mb_j= \f{\hbar}{P_0 \gamma} \f{|a_j|}{\gamma}, \;\;\;\; \mc_j=  \f{M}{m} \!\left(\! \f{\hbar}{P_0 \gamma} \!\right)^{\!\!2} \f{|a_2|}{\gamma} \f{t}{\tau_2} \, \omega(y_j)
\ea
and to use (\ref{ve}), (\ref{siga}), (\ref{mM}).

\n
We conclude this section noticing that our results (i.e. theorems 1, 2 in the next section) are still valid if the point interaction Hamiltonians are replaced by the Hamiltonians with a Coulomb potential $-e^2 |r- a_j|^{-1}$, as model Hamiltonians for the atoms. In fact, if we denote by $r_0$ the Bohr's radius, we can write the ground state energy and the corresponding eigenstate  as 

\ba\label{eidro}
&&\mathcal{E}_0=-\f{\hbar^2}{2mr_0^2}\\
&&\Xi_j (r) = \f{1}{r_0^{3/2}}\,
 \Xi^0 (r_0^{-1}(r-a_j)),  \;\;\;\;\;\;\; \Xi^0(x)= \f{1}{\sqrt{\pi}} \, e^{-|x|}
 \ea
Moreover the generalized eigenfunctions are 
\be
\Phi_j (r,k) =  \Phi^0 (r_0^{-1} ( r-a_j), r_0 k) , \;\;\;\;\;\;\; \left(\!-\Delta_x -\f{1}{|x|} \! \right)\! \Phi^0 (x,y) = y^2 \,\Phi^0 (x,y)   
\ee
It is now sufficient to replace $\gamma$, $E_0$, $\zeta_j$, $\phi_j$ by $r_0$, $\mathcal{E}_0$, $\Xi_j$, $\Phi_j$ and it is easily checked that the proofs work exactly in the same way.

\vs
\vs
\section{result}
\vs

\n
We are now in a position to formulate our result. 
We are interested in the computation of the probability that both atoms are ionized at time $t>0$. An exact  computation obviously requires the complete knowledge of the state $\Psi(t)$ of the system, which is not available. Following the original strategy of Mott  we shall limit to consider the second order approximation $\Psi_2(t)$  of the state $\Psi(t)$ which, iterating twice Duhamel's formula, is given by

\ba\label{psi2}
&&\Psi_2(t) = e^{-i \f{i}{\hbar} t H_0} \hat{\Psi}_{2}(t) \\
&&\hat{\Psi}_{2}(t)= \Psi_0 -i\f{\lambda}{\hbar} \int_0^{t} \!\! dt_1 \; e^{\f{i}{\hbar} t_1 H_0} H_1 e^{- \f{i}{\hbar} t_1 H_0}  \Psi_0 \nonumber\\
&&- \f{\lambda^2}{\hbar^2} 
\int_0^{t} \!\! dt_1 \; e^{\f{i}{\hbar} t_1 H_0} H_1 e^{- \f{i}{\hbar} t_1 H_0}
\int_0^{t_1} \!\! dt_2 \; e^{\f{i}{\hbar} t_2 H_0} H_1 e^{- \f{i}{\hbar} t_2 H_0}
\Psi_0 \label{hpsi2}
\ea

\n
Therefore we shall study the probability that both atoms are ionized up to second order in perturbation theory, i.e.

\be\label{prio}
\mathcal{P} (t) = \int \!\! dR \, dk_1 \, dk_2 \left| \int \!\! dr_1\, dr_2 \, \overline{\phi_1}(r_1,k_1) \overline{\phi_2}(r_2,k_2) \hat{\Psi}_2(R,r_1, r_2,t) \right|^2
\ee

\n
Our main result  is the characterization of the ionization probability $\mathcal{P}(t)$ for a fixed time $t>\tau_2$ and   it is summarized in theorems 1, 2 below.  
In theorem 1 we consider the case in which $a_2$ is not aligned with $a_1$ and the origin and we show that the ionization probability decays faster than  any power of $\ve$.

\vs
\n
{\bf Theorem 1.} {\em Let us fix $t>\tau_2$, $|a_1|<|a_2|$, $\hat{a}_1 \cdot \hat{a}_2 <1$ and let us assume} (\ref{ve}), (\ref{siga}), (\ref{mM}), $V \in \mathcal{S}(\erre^3)$. {\em Then for any $k \in \natu$ there exists $\mathcal{C}_k >0$ such that
\be\label{te1}
\mathcal{P}(t) \leq \left(\! \f{\lambda t}{\hbar}\!\right)^{\!4} \mathcal{N}^{2}\, \mathcal{C}_k \|\widetilde{V}\|^{4}_{W^{1,k}_{k}} \left[ \! \left(\! 1 - \f{|a_1|}{|a_2|} \!\right)^{\! \!-2k} + \left( 1 - \hat{a}_1 \! \cdot \! \hat{a}_2  \right)^{ -k}  \right] \ve^{2k-2}
\ee
}

\vs
\n
The constant $\mathcal{C}_k$ in (\ref{te1}), which will be specified during the proof, remains finite and strictly positive also if we take $|a_1|=|a_2|$ or $\hat{a}_1 \cdot \hat{a}_2=1$. In particular this means that  the estimate (\ref{te1}) is still meaningful if the angle between $a_1$ and $a_2$ is $O(\ve^p)$ with $0<p<1$, while it clearly fails for $p \geq 1$.      In the latter  case the second atom lies inside a small cone of aperture $O(\ve)$,  apex in the position $a_1$ of the first atom and axis on  the line joining the origin and $a_1$. This situation is considered  in theorem 2 where we compute the leading term of the asymptotic expansion for $\ve \rightarrow 0$ of the ionization probability.

\vs
\n
{\bf Theorem 2. } {\em Let us fix $t>\tau_2$, $|a_1|<|a_2|$,  $\hat{a}_1 \cdot  \hat{a}_2= \cos \chi_{\ve}$, where $\chi_{\ve} \in [0,\chi_0 \ve]$, $\chi_0 >0$, and let us assume} (\ref{ve}), (\ref{siga}), (\ref{mM}), $V \in \mathcal{S}(\erre^3)$. {\em Then, at the leading order for $\ve \rightarrow 0$,  we have

\ba\label{te2}
&&\mathcal{P}(t)  \sim \ve^2  ( \ve^{-1} \lambda_0)^{4} \left(\! \ve^{-1}\f{\gamma}{ |a_1|}  \!  \right)^{\!\! 4} \mathcal{N}^{2} \! \int_{\erre^9} \!\!\! dx dy dz \left| \int_{\erre^2} \!\!\! d\eta_1 d\eta_2 \, F(\eta_1, \eta_2;x,y,z) \right|^2
\ea
where the function $F$ is independent of $\ve$ and will be specified during the proof.

}

\vs

\n
In the following remarks we briefly comment on the above results.

\vs
\n
{\bf Remark 3.1.} The estimate  (\ref{te1}) is valid for $t$ larger but of the same order of $\tau_2$, while it loses its meaning for $t \rightarrow \infty$. This is only due to the method we use for the proof, based on second order perturbation  theory. A non-perturbative approach or a more detailed perturbative analysis should provide  an estimate which is uniform in time.    
We also remark that in theorem 2 we limit ourselves to the computation of the leading term, without making any attempt to estimate the remainder.  Such leading term is small for $\ve \rightarrow 0$ being proportional, as expected, to the solid angle that the atoms subtend at the origin.   
It would be interesting to extend the results given here at any order in perturbation theory and in particular to verify that the leading term has the same behavior for $\ve \rightarrow 0$ at all orders.

\vs
\n
{\bf Remark 3.2}. The results in theorems 1, 2 can be understood on the basis of the original physical argument given by Mott, which can be described as follows. At time zero the spherical wave starts to propagate in the chamber and at time $\tau_1$ it interacts with the atom in $a_1$. If, as result of the interaction, such atom is ionized then a localized wave packet emerges from $a_1$ with momentum along the direction $\overline{Oa_1}$. In order to obtain also ionization of the atom in $a_2$ the localized wave packet must hit the atom in $a_2$ (at time $\tau_2$) and this can happen only if $a_2$ approximately lies on the line $\overline{Oa_1}$. It should be stressed that such physical behavior is far from being universal and it strongly depends on our assumptions on the physical parameters of the model.

\vs
\n
{\bf Remark 3.3.} Finally we observe that our result 
 states that one can only observe straight tracks in a cloud chamber. 
With this we do not mean that there is any focusing of the support of the wave packet  of the  $\alpha$-particle along a classical straight trajectory, corresponding to the observed track. In fact the solution of the Schr\"odinger  equation with Hamiltonian (\ref{H}) and   initial datum (\ref{in}) has the form
\ba
&&\Psi (R,r_1,r_2,t)= \mathcal{F}_{00}(R,t) \zeta_1(r_1) \zeta_2(r_2) +\!\! \int\!\!dk_1 \, \mathcal{F}_{c0}(R, \!k_1,\!t) \phi_1(r_1,\!k_1) \,\cdot \zeta_2(r_2) \nonumber\\
&&+\!\! \int\!\!dk_2 \, \mathcal{F}_{0c}(R,\! k_2,\!t) \phi_2(r_2,\!k_2) \cdot \zeta_1(r_1) + \!\! \int\!\!dk_1 dk_2 \, \mathcal{F}_{cc}(R, \!k_1,\!k_2,\!t) \phi_1(r_1,\!k_1) \phi_2(r_2,\!k_2)\label{rp}
\ea
where the four  probability amplitudes $\mathcal{F}_{00}, \mathcal{F}_{c0}, \mathcal{F}_{0c}, \mathcal{F}_{cc}$  are localized  in different regions of the configuration space of the whole system and therefore  describe not interfering "quantum histories". If one interprets double ionization as the only case of macroscopic ionization, giving then rise to an observable track, one can say, in a pictorial language, that is along the track the expected value of the position of the $ \alpha $ particle in the states in which the track (as an observable) has expectation one.

\vs

\n
Let us outline the strategy of the proof of theorems 1, 2. 
The starting point is the following, more convenient,  representation formula for the ionization probability 

\ba\label{p2}
&&\mathcal{P}(t)=  \f{\lambda^4 t^4}{\hbar^4} \f{\mathcal{N}^2}{\ve^2} \int \!\! dx dy_1  dy_2 \left| \mathcal{G}_{12}^{\ve}(x,y_1,y_2,t) + \mathcal{G}_{21}^{\ve} (x,y_1,y_2,t) \right|^2
\ea

\n
 where for $l,j =1,2$, $j \neq l$ one has
\ba\label{gjl}
&&\mathcal{G}_{lj}^{\ve} ( x, y_1, y_2, t) \!=\!\! \int_{S^2} \!\!\!d \hat{u} \!\! \int_0^1 \!\!\!\! d \alpha \!\! \int_0^{\alpha} \!\!\!\!\! d \beta \!\!\int \!\! d \eta d \xi \,  
G_{lj}(\alpha, \beta, \eta, \xi ; x,y_1,y_2,t) \, e^{\f{i}{\ve} \Theta_{lj} (\hat{u}, \alpha, \beta, \eta, \xi ; x,y_1,y_2,t)}
\ea

\n
and dropping the parametric dependence on $x,y_1,y_2,t$

\ba\label{tejl}
&&\Theta_{lj} (\hat{u}, \alpha, \beta,\eta,\xi)=  \hat{u} \cdot \left(  x+ \mathfrak{a} (\alpha \eta + \beta \xi) \right) - \mathfrak{b}_j \hat{a}_j \cdot \eta - \mathfrak{b}_l \hat{a}_l \cdot \xi + \mathfrak{c}_j \alpha + \mathfrak{c}_l \beta \\
&&\nonumber\\
&&\label{gjl2}
G_{lj}(\alpha, \beta,\eta,\xi)= g(\eta,y_j)g(\xi,y_l) f(x+ \mathfrak{a}(\alpha \eta + \beta \xi)) e^{i \phi (\alpha,\beta,\eta, \xi)} \\
&&\nonumber\\
&&\label{phi}
\phi(\alpha,\beta,\eta,\xi)=  x \cdot (\eta +  \xi  ) +
 \f{\mathfrak{a}}{2} \,( \alpha \eta^2 + \beta \xi^2 + 2 \alpha \, \eta \cdot \xi )
\ea
 
\n
The proof of (\ref{p2}) is a long but straightforward computation and it is postponed to   the appendix (section 7). 

\n
Due to formula (\ref{p2}), we are reduced to the analysis of the two oscillatory integrals $ \mathcal{G}_{lj}^{\ve}$ corresponding to the possible graphs in the second order perturbative expansion. In particular $\mathcal{G}_{12}^{\ve}$ describes the graph in which the atom in $a_1$ is ionized before the atom in $a_2$ and $\mathcal{G}_{21}^{\ve}$ the opposite case. Since we always assume $|a_1|<|a_2|$,  we expect that the contribution of $\mathcal{G}_{21}^{\ve}$ is negligible. In fact, 
 in  section 4 we shall see  that the phase $\Theta_{21}$ has no critical points and then, by standard integration by parts, we shall prove that the contribution of the oscillatory integral $\mathcal{G}_{21}^{\ve}$ is $O(\ve^k)$, for any $k \in \natu$.

\n
The estimate of the term $\mathcal{G}_{12}^{\ve}$ is more delicate and we have to distinguish the non aligned   and the aligned case. It turns out that the phase $\Theta_{12}$ has no critical points in the first case and then the contribution of $\mathcal{G}_{12}^{\ve}$ is $O(\ve^k)$, for any $k \in \natu$. This will be proved in section 5, concluding also the proof of theorem 1.

\n
In section 6 we consider  the aligned case, where  the phase $\Theta_{12}$ has  a manifold of  critical points  parametrized by a vector in $\erre^2$. By a careful application of the stationary phase method to $\mathcal{G}_{12}^{\ve}$, we compute the leading term of the asymptotic expansion for $\ve \rightarrow 0$ and then we also conclude the proof of theorem 2.


\vs
\vs

\section{Estimate of $\mathcal{G}_{21}^{\ve}$}

\vs
\n
In this section we shall prove that the the contribution of the oscillatory integral $\mathcal{G}_{21}^{\ve}$ is negligible for any orientation of the unit vectors $\hat{a}_1$, $\hat{a}_2$.

\vs
\n
{\bf Proposition 4.1}.  {\em Let us fix $t>\tau_2$, $|a_1|<|a_2|$  and let us assume} (\ref{ve}), (\ref{siga}), (\ref{mM}), $V \in \mathcal{S}(\erre^3)$. {\em Then for any $k \in \natu$ there exists $\mathcal{C}_k >0$ such that

\be\label{SG21}
\int \!\! dx dy_1  dy_2 \left|  \mathcal{G}_{21}^{\ve} (x,y_1,y_2) \right|^2  \leq \mathcal{C}_k \, \|\widetilde{V}\|_{W^{1,k}_{k}}^{4}    \left(\! 1 - \f{|a_1|}{|a_2|} \!\right)^{\! \!-2k}   \ve^{2k}
\ee
}

\vs
\n
{\bf Proof.}  The crucial point is that the gradient of the phase

\be
\Theta_{21} = \hat{u} \cdot (x + \ma (\alpha \eta + \beta \xi)) - \mb_1 \hat{a}_1 \cdot \eta - \mb_2 \hat{a}_2 \cdot \xi + \mc_1 \alpha + \mc_2 \beta
\ee
doesn't vanish in the integration region. To see this it is sufficient to compute 

\ba
&&\sum_{k=1}^{3} \left( \f{\partial \Theta_{21}}{\partial \eta_k} \right)^2 +  \left( \f{\partial \Theta_{21}}{\partial \xi_k} \right)^2 = (\ma \alpha \hat{u} - \mb_1 \hat{a}_1 )^2 + (\ma \beta \hat{u} - \mb_2 \hat{a}_2 )^2 \nonumber\\
&& \geq (\ma \alpha  - \mb_1 )^2 + (\ma \beta - \mb_2 )^2 \equiv \ma^2 \left[ \left( \alpha - \f{\tau_1}{t} \right)^2 + \left( \beta - \f{\tau_2}{t} \right)^2 \right]
\ea

\n
In the region $\{ (\alpha,\beta) \in \erre^2 \;|\; 0\leq \alpha \leq 1, \; 0 \leq \beta \leq \alpha \}$ the r.h.s. of (\ref{gt21}) takes its minimum in $(\alpha_0 , \beta_0)$, with $\alpha_0 = \beta_0= \f{\tau_1 + \tau_2}{2t}  $, then

\be\label{gt21}
\sum_{k=1}^{3} \left( \f{\partial \Theta_{21}}{\partial \eta_k} \right)^2 +  \left( \f{\partial \Theta_{21}}{\partial \xi_k} \right)^2 \geq \Delta_{21}^2
\ee

\n
where
\be\label{de21}
\Delta_{21}=\f{\hbar}{\sqrt{2} M \gamma \sigma}(\tau_2 - \tau_1) \equiv \f{\hbar}{\sqrt{2} P_0 \gamma} \f{|a_2|}{\gamma} \left( 1 - \f{|a_1|}{|a_2|} \right)
\ee

\n
Notice that, under the assumptions (\ref{ve}), (\ref{siga}), (\ref{mM}) and $|a_1|<|a_2|$, $\Delta_{21}$ remains strictly larger than zero for any $\ve >0$. The estimate (\ref{gt21}) allows to control $\mathcal{G}_{21}$ using  standard non stationary phase methods (\cite{f}, \cite{ho}, \cite{bh}). In fact, recalling the identity 
\be
a\,e^{ib}= -i \, div  \left( e^{ib} \f{\nabla b}{|\nabla b|^2} \, a\right) +  i \, e^{ib} \, div \left( \f{\nabla b}{|\nabla b|^2} \,a \right)
\ee
and performing $k$ integration by parts we have
\be\label{intpar}
\int \! \! d\eta d \xi \,  G_{21} \, e^{\f{i}{\ve} \Theta_{21}} = (i \ve)^k \! \int \!\! d\eta d \xi \,   (L^k G_{21}) \, e^{\f{i}{\ve} \Theta_{21}}
\ee

\n
where the operator $L$ acts on the variables $\zeta = (\zeta_1, \ldots , \zeta_6) \equiv (\eta_1, \eta_2, \eta_3, \xi_1, \xi_2, \xi_3)$ as follows

\be\label{L}
L G_{21} = \sum_{j=1}^6 u_j \f{\partial G_{21}}{\partial \zeta_j} , \;\;\;\;\;\; u_j = \f{1}{|\nabla_{\zeta} \theta_{21}|^2} \f{\partial \theta_{21}}{\partial \zeta_j}
\ee

\n
and moreover

\be\label{Lk}
L^k G_{21} = \sum_{j_1 ... j_k =1}^6 u_{j_1} ...  u_{j_k} D^k_{\zeta_{j_1}...\zeta_{j_k}} G_{21}
\ee

\n
In (\ref{Lk}) we have denoted by $D^k_{\zeta_{j_1}...\zeta_{j_k}}$ the derivative of order $k$ with respect to $\zeta_{j_1}...\zeta_{j_k}$. 
From (\ref{Lk}), (\ref{L}), (\ref{gt21}), (\ref{de21}) we easily get the estimate

\be\label{stg21}
\left| \mathcal{G}_{21}^{\ve} \right| \leq \ve^k \!\! \int_{S^2} \!\!\! d \hat{u}\!\! \int_0^1 \!\! \!d \alpha \!\! \int_0^{\alpha} \!\!\! d \beta \!\! \int \!\! d \eta d \xi \; \left| L^k G_{21} \right| \leq 4 \pi \f{\ve^k}{\Delta_{21}^k} \int_0^1 \!\! \!d \alpha \!\! \int_0^{\alpha} \!\!\! d \beta \!\! \int \!\! d \eta d \xi \; \left| \sum_{j_1 ... j_k =1}^6 \!\!  D^k_{\zeta_{j_1}...\zeta_{j_k}} G_{21} \right|
\ee 

\n
If we square (\ref{stg21}), integrate w.r.t. the variables $x,y_1,y_2$ and use Schwartz  inequality we find

\be\label{L2g21}
\int \! dx dy_1 dy_2 \, |\mathcal{G}_{21}^{\ve}|^2 \leq 4 \pi^2 \; \f{\ve^{2k}}{\Delta_{21}^{2k}}\;  \sup_{\alpha, \beta} \left[ \int \! d\eta d \xi \left( \int \! dx dy_1 dy_2 \, \left| \sum_{j_1 ... j_k =1}^6 \!\!  D^k_{\zeta_{j_1}...\zeta_{j_k}} G_{21} \right|^2
 \right)^{\!1/2} \right]^2
\ee

\n
From the definition of $G_{21}$ (see (\ref{gjl2})), we have

\ba\label{deG21}
&&\sum_{j_1 ... j_k =1}^6 \!\!  | D^k_{\zeta_{j_1}...\zeta_{j_k}} G_{21}| 
\leq \mathcal{C}_k \sum_{i_1 = 1}^{k} | D^{i_1}_{\eta} g(\eta, y_1) | \sum_{i_2 =1}^{k}  | D^{i_2}_{\xi} g(\xi, y_2)| \sum_{i_3 =1}^k  
|D^{i_3}_{x}  f (x+ \ma (\alpha \eta + \beta \xi))| \nonumber\\
&& \cdot  \sum_{i_4 =1}^{k}  (|x| + \ma |\eta| + \ma |\xi| )^{i_4}
\ea

\n
The last term in (\ref{deG21}) can be easily estimated as follows

\ba\label{xz}
&&\sum_{i_4 =1}^{k} (|x| + \ma |\eta| + \ma |\xi| )^{i_4} \leq 
\sum_{i_4 =1}^{k} (|x + \ma (\alpha \eta + \beta \xi)| + 2 \ma |\eta| + 2 \ma |\xi| )^{i_4} \nonumber\\
&&\leq \sum_{i_4 =1}^{k}   \left( \sqrt{2}   \,4 \ma^2  <\! x + \ma (\alpha \eta + \beta \xi) \!> <\! \eta \!> <\! \xi \!> \right)^{i_4} \nonumber\\
&&\leq \left(\!  \sum_{i_4 =1}^{k}  \ma^{2 i_4} 2^{5 i_4/ 2}  \! \right) <\! x + \ma (\alpha \eta + \beta \xi) \!> ^k <\! \eta \!>^k  <\! \xi \!>^k
\ea

\n
where $< \! x \! >^2 = 1 +x^2$, $x \in \erre^3$. Hence

\ba\label{xzbis}
&&\sum_{j_1 ... j_k =1}^6 \!\!  | D^k_{\zeta_{j_1}...\zeta_{j_k}} G_{21}| \leq \mathcal{C}_k  <\! \eta \!>^k \sum_{i_1 = 1}^{k} | D^{i_1}_{\eta} g(\eta, y_1)| <\! \xi \!>^k  \sum_{i_2 =1}^{k}  | D^{i_2}_{\xi} g(\xi, y_2)|   \nonumber\\
&&\cdot  <\! x + \ma (\alpha \eta + \beta \xi) \!> ^k  \sum_{i_3 =1}^k  
|D^{i_3}_{x}  f (x+ \ma (\alpha \eta + \beta \xi))|
\ea

\n
Moreover, recalling the definition (\ref{g}), we  get

\ba\label{xzter}
&&\sum_{j_1 ... j_k =1}^6 \!\!  | D^k_{\zeta_{j_1}...\zeta_{j_k}} G_{21}| \leq \mathcal{C}_k  <\! \eta \!>^k \sum_{i_1 = 1}^{k} | D^{i_1}_{\eta} \widetilde{V}(\eta)| <\! \xi \!>^k  \sum_{i_2 =1}^{k}  | D^{i_2}_{\xi} \widetilde{V}(\xi)|   \nonumber\\
&&\cdot \! <\! x \! +\! \ma (\alpha \eta + \beta \xi) \!> ^k \! \sum_{i_3 =1}^k  
|D^{i_3}_{x}  f (x \!+\! \ma (\alpha \eta + \beta \xi))|
\sum_{i_4 =1}^{k } |D^{i_4}_{\eta} h(\eta,y_1)| \sum_{i_5 =1}^{k } |D^{i_5}_{\xi} h(\xi,y_2)| \nonumber\\
&&
\ea

\n
Using (\ref{xzter}) in estimate (\ref{L2g21}) we find

\ba\label{L2g21bis}
&&\int \! dx dy_1 dy_2 \, |\mathcal{G}_{21}^{\ve}|^2 
\leq \f{\ve^{2k}}{\Delta_{21}^{2k}}\,  \mathcal{C}_k \Bigg\{  \int \!\! d \eta  <\! \eta \!>^k \! \sum_{i_1 =1}^{k} |D^{i_1}_{\eta} \widetilde{V}(\eta)| 
\int \!\! d \xi  <\! \xi \!>^k \! \sum_{i_2 =1}^{k} |D^{i_2}_{\xi} \widetilde{V}(\xi)|  \nonumber\\
&& \cdot
\Bigg[ \int \!\! dy_1 \!\!\; \left( \sum_{i_3=1}^{k} |D^{i_3}_{\eta} h(\eta,y_1)| \right)^{\!\!2}
\int \!\! dy_2 \!\!\; \left( \sum_{i_4=1}^{k} |D^{i_4}_{\xi} h(\xi,y_2)| \right)^{\!\!2}
\Bigg]^{\! \!1/2}
\Bigg\}^{\!\!2} \nonumber\\
&&\leq \f{\ve^{2k}}{\Delta_{21}^{2k}}\,  \mathcal{C}_k \, \|\widetilde{V}\|_{W^{1,k}_{k}}^{4} \Bigg[ \sup_{\eta} \int \!\! dy \!\!\; \left( \sum_{m=1}^{k} |D^{m}_{\eta} h(\eta,y)| \right)^{\!\!2} \Bigg]^{\!2} \nonumber\\
&&\leq \f{\ve^{2k}}{\Delta_{21}^{2k}}\,  \mathcal{C}_k \, \|\widetilde{V}\|_{W^{1,k}_{k}}^{4} \Bigg[ 
\sum_{m=1}^k \sup_{\eta} \Big( \int \!\! dy \, |D^m_{\eta} h(\eta,y)|^2 \Big)^{\!1/2} \Bigg]^{\!4}
\ea

\n
It remains  to show that the last term in the r.h.s. of (\ref{L2g21bis}) is finite.  From the definition (\ref{h}) we have 


\n

\be\label{Dh}
D^m_{\eta} h (\eta,y) = \f{(-i)^m}{(2 \pi)^{3/2}} \int\!\! dx \; e^{-i \eta \cdot x} \; x_1^{m_1} x_{2}^{m_2} x_{3}^{m_3} \; \zeta_0(x) \,\overline{\phi^0}(x,y)
\ee

\n
where $m_1+m_2+m_3=m$.  We recall that the integral kernel $\overline{\phi^0}(x,y) $ defines a bounded operator in $L^2(\erre^3)$, with norm less or equal to one. Hence 

\ba\label{Dh1}
&&\int \!\! dy \, |D^m_{\eta} h(\eta,y)|^2  \leq \f{1}{(2 \pi)^{3}}  \int\!\! dx \; \left| x_1^{m_1} x_{2}^{m_2} x_{3}^{m_3} \; \zeta_0(x) \right|^2 < \infty
\ea

\n
Taking into account  inequality (\ref{Dh1}) in (\ref{L2g21bis}), we conclude the proof.

\vs
\hfill $\Box$


\vs
\vs

\section{Estimate of $\mathcal{G}_{12}^{\ve}$ in the case $\hat{a}_1 \cdot \hat{a}_2 < 1$}

\vs
\n
Following the same line of the previous section, we prove that the contribution of $\mathcal{G}_{12}^{\ve}$ to the ionization probability is negligible provided that the two unit vectors $\hat{a}_1$ and $\hat{a}_2$ are not parallel. Of course, the estimate shall crucially depend on the angle between the two unit vectors.

\vs
\n
{\bf Proposition 5.1.} {\em  Let us fix $t>\tau_2$,   $\hat{a}_1\! \cdot \! \hat{a}_2 <1$ and let us assume} (\ref{ve}), (\ref{siga}), (\ref{mM}), $V \in \mathcal{S}(\erre^3)$. {\em Then for any $k \in \natu$ there exists $\mathcal{C}_k >0$ such that

\be\label{SG12}
\int \!\! dx dy_1  dy_2 \left|  \mathcal{G}_{12}^{\ve} (x,y_1,y_2) \right|^2  \leq \mathcal{C}_k \, \|\widetilde{V}\|_{W^{1,k}_{k}}^{4}   \big(1- \hat{a}_1 \! \cdot \! \hat{a}_2 \big)^{-k}  \, \ve^{2k}
\ee

}

\vs
\n
{\bf Proof.}   As in the case of proposition 4.1, we consider the phase

\be\label{te12}
\Theta_{12}= \hat{u} \cdot (x + \ma (\alpha \eta + \beta \xi)) - \mb_2 \hat{a}_2 \cdot \eta - \mb_1 \hat{a}_1 \cdot \xi + \mc_2 \alpha + \mc_1 \beta
\ee

\n
and we show that its gradient is strictly different from zero in the integration region. In fact

\be\label{gte12}
\sum_{k=1}^{3} \left( \f{\partial \Theta_{12}}{\partial \eta_k} \right)^2 +  \left( \f{\partial \Theta_{12}}{\partial \xi_k} \right)^2 = (\ma \alpha \hat{u} - \mb_2 \hat{a}_2 )^2 + (\ma \beta \hat{u} - \mb_1 \hat{a}_1 )^2 
\ee

\n
The r.h.s. of (\ref{gte12}), considered as  a function of  the variables $(\alpha, \beta)$, takes its minimum in $(\alpha_1, \beta_1)= (\f{\mb_2}{\ma}\, \hat{u} \cdot \hat{a}_2, \f{\mb_1}{\ma}\, \hat{u} \cdot \hat{a}_1)$. Hence

\ba\label{gte12bis}
&&\sum_{k=1}^{3} \left( \f{\partial \Theta_{12}}{\partial \eta_k} \right)^2 +  \left( \f{\partial \Theta_{12}}{\partial \xi_k} \right)^2  \geq \mb_2^2 \,  (\hat{u} \cdot \hat{a}_2 \, \hat{u} - \hat{a}_2)^2 + \mb_1^2 \,  (\hat{u} \cdot \hat{a}_1 \, \hat{u} - \hat{a}_1)^2 \nonumber\\
&&\geq \min \{\mb_1^2, \mb_2^2\}  \Big[ 2 - (\hat{u} \cdot \hat{a}_2)^2 - (\hat{u} \cdot \hat{a}_1)^2 \Big]
\ea

\n
Let us fix a  frame of reference such that  $\hat{a}_1 =(0,0,1)$, $\hat{a}_2=( \sin \chi, 0, \cos \chi)$, with $\chi \in (0, \pi]$. Then the r.h.s. of (\ref{gte12bis}), considered as a function of $\hat{u}=( \sin \theta \cos \phi, \sin \theta \sin \phi, \cos \theta)$, $ \theta \in [0,\pi]$, $\phi \in [0,2 \pi)$, takes its minimum in $\hat{u}_0=(\sin \f{\chi}{2}, 0, \cos \f{\chi}{2})$. Hence

\be\label{gte12ter}
\sum_{k=1}^{3} \left( \f{\partial \Theta_{12}}{\partial \eta_k} \right)^2 +  \left( \f{\partial \Theta_{12}}{\partial \xi_k} \right)^2  \geq  \Delta_{12}^2                                                  \ee

\n
where

\be\label{de12}
\Delta_{12} = \sqrt{2} \, \min \{\mb_1, \mb_2\} \sqrt{1 - \cos^2 \f{\chi}{2}}  = \f{\hbar}{P_0 \gamma} \f{ \min \{|a_1|, |a_2|\} }{\gamma}  \sqrt{1 - \hat{a}_1 \cdot \hat{a}_2}
\ee

\n
We notice that, under the assumptions (\ref{ve}), (\ref{siga}), (\ref{mM}) and $\hat{a}_1 \cdot \hat{a}_2 <1$, $\delta_{12}$ remains strictly larger than zero for any $\ve >0$.

\n
From now on the proof proceeds exactly in the same way as in the previous proposition 4.1 and we omit the details.

\vs
\hfill $\Box$

\vs
\n
{\bf Proof of theorem 1}. From (\ref{p2}) and taking into account propositions 4.1, 5.1 we immediately get the proof.

\vs
\hfill$\Box$


\vs
\vs

\section{the stationary case}

\vs
\n
Here we consider the case  $\hat{a}_1 \! \cdot  \hat{a}_2 = 1 - O(\ve^q)$, with $q \geq 2$. We shall see that  the phase  of the oscillatory integral $\mathcal{G}_{12}^{\ve}$ has stationary points when exactly this  case occurs. This implies that the ionization probability $\mathcal{P}(t)$ is not negligible for $\ve$ small as in the previous situation and  the  leading term of its asymptotic expansion in powers of $\ve$ can be computed.

\n
Throughout this section we shall fix the unit vectors $\hat{a}_1$, $\hat{a}_2$ as follows

\be\label{a12al}
\hat{a}_1 = (0,0,1), \;\;\;\;\;\; \hat{a}_2 = (\sin \chi_{\ve}, 0,  \cos \chi_{\ve})
\ee
where $\chi_{\ve} \in[0,\chi_0\, \ve]$, $\chi_0>0$.   Moreover, in order to characterize the asymptotic behavior of $\mathcal{G}_{12}^{\ve}$, we introduce a convenient decomposition of the unit sphere $S^2$.  
More precisely we define $\Gamma_{\bar{\theta}}$ as the portion of $S^2$ inside a cone with apex in the origin, axis parallel to $\hat{a}_1$,  aperture $\bar{\theta}$, $0<\bar{\theta}<\f{\pi}{2}$, and we denote   
$ \bar{\Gamma}_{\bar{\theta}} = S^2 \setminus  \Gamma_{\bar{\theta}}$. For any choice of $\hat{a}_1$, $\hat{a}_2$ as in (\ref{a12al}) the corresponding decomposition of $\mathcal{G}_{12}^{\ve}$ reads

\ba\label{gns}
&&\mathcal{G}_{12}^{\ve}=
 \mathcal{G}_{12}^{\ve,n}  +  \mathcal{G}_{12}^{\ve,s}\\
&&\mathcal{G}_{12}^{\ve,n} =   \int_{\bar{\Gamma}_{\bar{\theta}}} \!\!\!\! d \hat{u} \! \int_0^1 \!\!\! d\alpha \! \int_{0}^{\alpha} \!\!\!\!d \beta \!\int \!\! d\eta d\xi \,G_{12}^{\ve} \; e^{\f{i}{\ve} \Theta} \label{gn}\\
&&\mathcal{G}_{12}^{\ve,s} =  \int_{\Gamma_{\bar{\theta}}} \!\!\!\! d \hat{u} \! \int_0^1 \!\!\! d\alpha \! \int_{0}^{\alpha} \!\!\!\!d \beta \!\int \!\! d\eta d\xi \,G_{12}^{\ve} \; e^{\f{i}{\ve} \Theta}
 \label{gs}
\ea 
where 
\ba\label{gve1}
&&G_{12}^{\ve}=  G_{12} \, e^{i \delta_{\ve}} \\
&&\delta_{\ve} =- \f{\sin \chi_{\ve}}{\ve}\, \mb_2 \eta_1 + \f{1- \cos \chi_{\ve}}{\ve} \,\mb_2 \eta_3 \\
&&\Theta = \hat{u} \cdot (x+\ma (\alpha \eta + \beta \xi)) - \mb_1 \xi_3 - \mb_2 \eta_3 +\mc_2 \alpha +\mc_1 \beta
\label{t1}
\ea

\n
We shall analyze the asymptotic behavior of the two oscillatory integrals $\mathcal{G}_{12}^{\ve,n}$ and $\mathcal{G}_{12}^{\ve,s}$ separately. We first show  that the phase $\Theta$ has no stationary points in $\bar{\Gamma}_{\bar{\theta}}$ and then the contribution of $\mathcal{G}_{12}^{\ve,n}$ is negligible.

\vs
\n
{\bf Proposition 6.1}.  {\em Let us fix $t> \tau_2$, $\hat{a}_1$, $\hat{a}_2$ as in}  (\ref{a12al}) {\em and let us assume} (\ref{ve}), (\ref{siga}), (\ref{mM}), $V \in \mathcal{S}(\erre^3)$. {\em Then for any $k \in \natu$ there exists $\mathcal{C}_k >0$ such that

\be\label{SGN12}
\int \!\! dx dy_1  dy_2 \left|  \mathcal{G}_{12}^{\ve,n} (x,y_1,y_2) \right|^2  \leq \mathcal{C}_k \, \|\widetilde{V} \|^{4}_{W_{k}^{1,k}} \left(\f{ \ve}{\sin \bar{\theta}} \right)^{2k} 
\ee

}

\vs
\n
{\bf Proof.} If we denote $\hat{u}=(\sin \theta \cos \phi, \sin \theta \sin \phi, \cos \theta) \in \bar{\Gamma}_{\bar{\theta}}$, we have

\ba\label{sTh1}
&&\sum_{k=1}^{3} \left( \f{\partial \Theta}{\partial \eta_k} \right)^2 +  \left( \f{\partial \Theta}{\partial \xi_k} \right)^2 = \ma^2 (\alpha^2 + \beta^2) + \mb_1^2 + \mb_2^2 -2 \ma (\mb_1 \beta + \mb_2 \alpha) \cos \theta\nonumber\\
&&\geq  \ma^2 (\alpha^2 + \beta^2) + \mb_1^2 + \mb_2^2 -2 \ma (\mb_1 \beta + \mb_2 \alpha) \cos \bar{\theta}
\ea

\n
The r.h.s. of (\ref{sTh1}) takes its minimum in $(\alpha_2, \beta_2)= \left( \f{\mb_2}{\ma} \cos \bar{\theta} ,  \f{\mb_1}{\ma} \cos \bar{\theta} \right)$, then 

\be\label{sTh11}
\sum_{k=1}^{3} \left( \f{\partial \Theta}{\partial \eta_k} \right)^2 +  \left( \f{\partial \Theta}{\partial \xi_k} \right)^2 \geq \Delta^2
\ee
where
\be\label{D1}
\Delta = \sqrt{2}\, \min \{\mb_1, \mb_2 \} \sin \bar{\theta} = \sqrt{2}\, \f{\hbar}{P_0 \gamma} \f{\min \{|a_1|, |a_2|\} }{\gamma} \sin \bar{\theta}
\ee

\n
Exploiting estimate (\ref{sTh11}), (\ref{D1}) it is now straightforward to obtain (\ref{SGN12}). 

\vs
\hfill $\Box$

\vs
\n
{\bf Remark 6.1}.   
We notice that the estimate (\ref{SGN12}) is still meaningful if we choose the angle $\bar{\theta}$ depending on $ \ve$ and such that  $\bar{\theta} = O(\ve^{d})$, with $0<d<1$.  This in particular means that only a small fraction of the unit sphere, of area $O(\ve^2)$ around the direction $\hat{a}_1$, can give a non trivial contribution to the ionization probability.

\vs

\n
Let us consider the oscillatory integral $\mathcal{G}_{12}^{\ve,s}$. It turns out that 
 the  phase $\Theta$ has a manifold of critical points in the integration region, parametrized by a vector in $\erre^2$. Therefore we fix the variables $(\eta_1, \eta_2) \in \erre^2$ as parameters and we write  $\mathcal{G}^{\ve,s}_{12}$ in the form
 
 \ba\label{Geta}
 &&\mathcal{G}^{\ve,s}_{12}=  \int \!\! d\eta_1 d\eta_2 \, \mathcal{I}^{\ve}(\eta_1, \eta_2) \\
 &&\mathcal{I}^{\ve}(\eta_1, \eta_2) =  \int_{\Omega} \!\!\!dq \, G^{\ve}_{12} (q; \eta_1, \eta_2) \, e^{\f{i}{\ve} \Theta (q;\eta_1,\eta_2)}
 \label{Ieta}
 \ea
 where
\be\label{q}
 \Omega= \!\{  q \equiv (\hat{u}, \alpha, \beta, \eta_3, \xi) \;|\; \hat{u} \in \Gamma_{\bar{\theta}}, \; \alpha \in [0,1], \; \beta \in [0,\alpha],\; \eta_3 \in \erre ,\; \xi \in \erre^3 \}
 \ee

\n
In the next lemma we show that for each  value of the parameters $(\eta_1,\eta_2)$ the phase in (\ref{Ieta}) has one, non degenerate   stationary point. It is relevant that the value of the phase  and of the hessian of the phase at the critical point do not depend on $(\eta_1, \eta_2)$.

\vs
\n
{\bf Lemma 6.2}. {\em For each $(\eta_1, \eta_2) \in \erre^2$ the phase $\Theta (q; \eta_1, \eta_2)$, $q\in \Omega$, has exactly one critical point

\ba\label{cp}
&&q_0\equiv (\hat{u}^0 \!, \, \alpha^0, \,\beta^0, \,\eta_3^0, \,\xi_1^0, \, \xi_2^0,\, \xi_3^0)
\ea
where
\ba
&&\hat{u}^0=(0,0,1), \; \;\;\;\; \alpha^0= \f{\mb_2}{\ma},\;\;\;\; \; \beta^0= \f{\mb_1}{\ma},\; \;\; \;\;\eta_3^0= -\f{\mc_2}{\ma},\label{ptc1}\\
&& \xi_1^0= -\f{x_1 + \mb_2 \eta_1}{\mb_1},\;\;\;\; \; \xi_2^0= -\f{x_2 + \mb_2 \eta_2}{\mb_1},\;\;\;\;\; \xi_3^0= - \f{\mc_1}{\ma} \label{ptc2}
\ea

\n
and moreover

\ba\label{T2cp}
&&\Theta^0 \equiv \Theta (q_0;  \eta_1, \eta_2)=x_3 +  \f{\mb_1 \mc_1}{\ma} +  \f{\mb_2 \mc_2}{\ma} 
\\
&&\label{DT2cp}
|D^2\Theta^0|\equiv|D^2_q \Theta(q_0;\eta_1, \eta_2)|= \ma^4 \mb_1^4
\ea

}

\vs
\n
{\bf Proof}. In order to compute the critical points of the phase (\ref{t1}) as a function of   $q \in \Omega$ it is convenient to write $\hat{u}=( \mu, \, \nu, \, \sqrt{1 - \mu^2 - \nu^2})$, where $(\mu, \nu) \in \erre^2$ with $\mu^2 + \nu^2 \leq \sin^2 \bar{\theta}$. Therefore 

\ba\label{t3}
&&\Theta(q;\eta_1,\eta_2)= \mu\,  w_1 +\nu \, w_2 +\sqrt{1-\mu^2 -\nu^2}\,  w_3 - \mb_2 \eta_3 - \mb_1 \xi_3 + \mc_2 \alpha +\mc_1 \beta
\ea
where we have introduced the short hand notation
\be\label{w_j}
w=(w_1, w_2,w_3), \;\;\;\;\;\;\;\; w_j= x_j + \ma (\alpha \eta_j + \beta \xi_j)
\ee

\n
By an explicit computation, one finds that  the critical points are solutions of the system

\ba
&&\f{\partial \Theta}{\partial \mu}= w_1 -\f{\mu w_3}{\sqrt{1-\mu^2 -\nu^2}}=0  \label{d1}\\
&&\f{\partial \Theta}{\partial \nu}  = w_2 - \f{\nu w_3}{\sqrt{1-\mu^2 -\nu^2}}=0 \label{d2}\\
&&\f{\partial \Theta}{\partial \alpha}  = \ma \mu  \eta_1 + \ma \nu  \eta_2 + \ma \sqrt{1-\mu^2 -\nu^2} \, \eta_3 + \mc_2 =0  \label{d3} \\
&&\f{\partial \Theta}{\partial \beta}  = \ma \mu \xi_1 + \ma \nu \xi_2 + \ma \sqrt{1-\mu^2 -\nu^2} \,  \xi_3 + \mc_1 =0 \label{d4} \\
&&\f{\partial \Theta}{\partial \eta_3}  = \ma \sqrt{1-\mu^2 -\nu^2} \, \alpha - \mb_2 =0   \label{d5}\\
&&\f{\partial \Theta}{\partial \xi_1} = \ma \mu \beta=0  \label{d6}  \\
&&\f{\partial \Theta}{\partial \xi_2}   = \ma \nu \beta =0  \label{d7}\\
&&\f{\partial \Theta}{\partial \xi_3}    = \ma \sqrt{1-\mu^2 -\nu^2}\,  \beta - \mb_1 =0 \label{d8}
\ea


\n
First we notice that $\alpha$ and $\beta$ cannot be zero,  otherwise from (\ref{d5}), (\ref{d8})  one would have $\mb_2= \mb_1 =0$. Then   from (\ref{d6}), (\ref{d7}) we have $\mu=\nu=0$ and from (\ref{d5}), (\ref{d6}) we have $\alpha = \f{\mb_2}{\ma}$, $\beta= \f{\mb_1}{\ma}$. Exploiting the remaining equations it is now trivial to find the unique solution (\ref{ptc1}), (\ref{ptc2}).
Furthermore the value of the phase at the critical point (\ref{T2cp})  is easily obtained. For the proof of (\ref{DT2cp})  we need the second derivatives of the phase evaluated at the critical point


\ba
&&\f{\partial^2 \Theta}{\partial \mu^2}= -w_3\,, \;\;\;\;  
\f{\partial^2 \Theta}{\partial \mu \partial \nu}= 0 \, , \;\;\;\; 
\f{\partial^2 \Theta}{\partial \mu \partial \alpha}= \ma \eta_1 \,, \;\;\;\;  
\f{\partial^2 \Theta}{\partial \mu \partial \beta}= \ma \xi_1 \,, \;\;\;\;
\f{\partial^2 \Theta}{\partial \mu \partial \eta_3}= 0 \nonumber\\ 
&&\f{\partial^2 \Theta}{\partial \mu \partial \xi_1}= \mb_1 \,, \;\;\;\;  
\f{\partial^2 \Theta}{\partial \mu \partial \xi_2}=0 \,,\;\;\;\;  
\f{\partial^2 \Theta}{\partial \mu \partial \xi_3}=0 \nonumber\\
&&\f{\partial^2 \Theta}{\partial \nu^2}=- w_3 \,,\;\;\;\; 
\f{\partial^2 \Theta}{\partial \nu \partial \alpha}=\ma \eta_2 \,, \;\;\;\; 
\f{\partial^2 \Theta}{\partial \nu \partial \beta}= \ma \xi_2 \,,\;\;\;\; 
\f{\partial^2 \Theta}{\partial \nu \partial \eta_3}=0\nonumber\\ 
&&\f{\partial^2 \Theta}{\partial \nu \partial \xi_1}=0 \,,\;\;\;\; 
\f{\partial^2 \Theta}{\partial \nu \partial \xi_2}=\mb_1\,,\;\;\;\;
\f{\partial^2 \Theta}{\partial \nu \partial \xi_3}= 0\nonumber
\ea
\ba
&&\!\!\!\!\!\!\!\!\!\! \f{\partial^2 \Theta}{\partial \alpha^2}= 0\,,\;\;\;\; 
\f{\partial^2 \Theta}{\partial \alpha \partial \beta}= 0\,,\;\;\;\;
\f{\partial^2 \Theta}{\partial \alpha \partial \eta_3}= \ma \,,\;\;\;\;
\f{\partial^2 \Theta}{\partial \alpha \partial \xi_j}=0  \nonumber\\
&&\!\!\!\!\!\!\!\! \!\!  \f{\partial^2 \Theta}{\partial \beta^2}= 0 \,,\;\;\;\; 
\f{\partial^2 \Theta}{\partial \beta \partial \eta_3}=0 \,,\;\;\;\; 
\f{\partial^2 \Theta}{\partial \beta \partial \xi_1}=0 \,,\;\;\;\; 
\f{\partial^2 \Theta}{\partial \beta \partial \xi_2}=0\,,\;\;\;\; 
\f{\partial^2 \Theta}{\partial \beta \partial \xi_3}=\ma \nonumber\\ 
&&\!\!\!\!\!\!\!\! \!\! \f{\partial^2 \Theta}{\partial \eta_3^2}=0 \,,\;\;\;\; 
\f{\partial^2 \Theta}{\partial \eta_3 \partial \xi_j}=0 \,,\;\;\;\; 
\f{\partial^2 \Theta}{\partial \xi_j \partial \xi_k}=0
\ea
\n
The computation of the hessian is now a tedious but straightforward exercise  and it is omitted for the sake of brevity.

\vs
\hfill $\Box$

\vs
\n
We are now ready to conclude the proof of theorem 2.

\vs
\n
{\bf Proof of theorem 2}. Exploiting the stationary phase theorem (\cite{f}, \cite{ho}, \cite{bh}) and the previous lemma, we find the leading term of the asymptotic expansion of (\ref{Ieta}) for $\ve \rightarrow 0$

\ba\label{Ieta0}
&&\mathcal{I}^{\ve}(\eta_1,\eta_2) \sim \f{(2\pi \ve)^4}{\ma^2 \mb_1^2}\; e^{\f{i}{\ve} \Theta^0} \,G_{12}(q_0;\eta_1,\eta_2)\, e^{i\delta_0} \, e^{i \f{\pi}{4}\mu_0}
\ea

\n
where $\mu_0$ denotes the signature of the hessian matrix at the critical point and moreover
\be
\delta_0 = - \lim_{\ve \rightarrow 0} \f{\sin \chi_{\ve}}{\ve} \,\, \mb_2 \eta_1
\ee
In particular $\delta_0 \neq 0$ if $\chi_{\ve} =O(\ve)$ and $\delta_0 =0$ if  $\chi_{\ve}=O(\ve^q)$, $q>1$, or $\chi_{\ve}=0$.  From (\ref{Geta}) we also obtain

\ba\label{Get}
\mathcal{G}_{12}^{\ve,s}
\sim \; \f{(2\pi\ve)^4}{\ma^2 \mb_1^2}   \, e^{\f{i}{\ve} \Theta_2^0} \! \!\int \!\!d\eta_1 d\eta_2 \, G_{12}(q_0;\eta_1,\eta_2) \, e^{i \delta_0}\, e^{ i\f{\pi}{4}\mu_0} 
\ea

\n
We notice that the integrand in (\ref{Get}) is a function of $x$ (position of the $\alpha$-particle), $y_j$ (momentum of the $j$-th ionized atom) and $\eta_1, \eta_2$. Hence we denote 
\ba\label{F}
&&F(\eta_1,\eta_2;x,y_1,y_2)\equiv  (2\pi)^4  \,  G_{12}(q_0;\eta_1,\eta_2) \, e^{ i \delta_0 }\, e^{ i\f{\pi}{4}\mu_0}  
\ea
and, taking into account (\ref{p2}), proposition 4.1, (\ref{gns}), proposition 6.2 and (\ref{Get}), we find 

\be\label{pas}
\mathcal{P}(t) \sim \left(\! \f{\lambda t}{\hbar} \! \right)^{\!4}  \f{\mathcal{N}^2}{\ma^4 \mb_1^4} \, \ve^6 \! \int \!\! dxdy dz \, \left|\int \!\! d\eta_1 d\eta_2 \, F(\eta_1,\eta_2 ; x,y,z) \right|^2
\ee
Using the definition of $\ma$, $\mb_1$, $\tau_1$ in (\ref{pas}),  we easily get formula (\ref{te2}). It remains to show that the integral in (\ref{pas}) is finite. From the definitions (\ref{gjl2}), (\ref{g}), (\ref{h}) and the boundedness of $h(\xi,y)$ we have

\be\label{F2}
|F(\eta_1,\eta_2;x,y,z)| \leq c \, |\widetilde{V}(\eta_1,\eta_2,\eta_3^0)| |\widetilde{V}(\xi_1^0,\xi_2^0,\xi_3^0)||f(w_0)|
\ee
where
\ba
&&\eta_3^0= -\ve \f{M}{m} \omega(z)\, , \;\;\;
\xi_1^0= -\ve^{-1} \f{\gamma}{|a_1|} x_1 - \f{\tau_2}{\tau_1} \eta_1\,, \;\;\;
\xi_2^0= -\ve^{-1} \f{\gamma}{|a_1|} x_2 - \f{\tau_2}{\tau_1} \eta_2\,, \nonumber\\
&& 
\xi_3^0= - \ve \f{M}{m} \omega(y)\,, \;\;\; w_0=x_3 - \ve^2 \f{M}{m} \f{|a_1|}{\gamma} \omega(y) - \ve^2 \f{M}{m} \f{|a_2|}{\gamma} \omega(z)\, 
\ea
and $\omega(y)$ is defined in (\ref{ome}). Then we write

\ba
&&\int \!\! dxdydz \, \left|\int \!\! d\eta_1 d\eta_2 \, F(\eta_1,\eta_2 ; x,y,z) \right|^2 \!\nonumber\\
&&\leq c \! \int \!\! dx_1 dx_2 dy dz \left(\! \int \!\! d\eta_1 d\eta_2  |\widetilde{V}(\eta_1,\eta_2,\eta_3^0)| |\widetilde{V}(\xi_1^0,\xi_2^0,\xi_3^0)| \!\right)^{\!2} \int \!\! dx_3 \, |f(w_0)|^2
\ea
Using the Schwartz inequality in the integral with respect to $(\eta_1,\eta_2)$ we have

\ba
&&\int \!\! dxdydz \, \left|\int \!\! d\eta_1 d\eta_2 \, F(\eta_1,\eta_2 ; x,y,z) \right|^2 \!\nonumber\\
&&\leq c \! \int \!\! dz \!\! \int\!\! d\eta_1 d\eta_2 <\! \eta_1 \!>^4 <\! \eta_2 \!>^4 \! |\widetilde{V}(\eta_1,\eta_2,\eta_3^0)|^2 \! \int\!\! dy \!\! \int\!\! dx_1 dx_2  |\widetilde{V}(\xi_1^0,\xi_2^0,\xi_3^0)|^2 \! \int \!\! dx_3 \, |f(w_0)|^2\nonumber\\
&&
\ea
The last integral is finite due to the assumptions on $V$ and this concludes the proof of the theorem.

\vs
\hfill $\Box$


\vs
\vs
\section{appendix}
\vs
 
\n
Here we give a proof of the representation formula (\ref{p2}). The relevant object to compute is the probability amplitude in (\ref{prio})

\ba\label{calf}
&&\mathcal{F}(R, k_1,k_2, t)= \int \!\! dr_1\, dr_2 \, \overline{\phi_1}(r_1,k_1) \overline{\phi_2}(r_2,k_2) \hat{\Psi}_2(R,r_1, r_2,t) 
\ea

\n
We notice that 

\ba\label{w12}
&& e^{\f{i}{\hbar} t H_0} H_1 e^{- \f{i}{\hbar} t H_0} = W_1 (t) + W_2(t)\\
&&W_j(t)= e^{\f{i}{\hbar} t K_0} e^{\f{i}{\hbar} t K_j} V_j 
e^{-\f{i}{\hbar} t K_0} e^{-\f{i}{\hbar} t K_j}
\ea

\n
where $V_j$ denotes the multiplication operator by

\be\label{vj}
V_j(R,r_j)=V(\gamma^{-1}(R-r_j))
\ee

\n
Let us rewrite the r.h.s. of (\ref{calf}) in a more convenient form. 
We observe that the operator $W_j(t)$ acts non trivially only on the variable $R$ and $r_j$. Exploiting this fact it is easily seen that

\ba\label{calfa}
&&\mathcal{F}= \mathcal{F}_{12} + \mathcal{F}_{21}\\
&&\mathcal{F}_{lj}(R, \!k_1,\!k_2,\! t)
\!=\! \f{\lambda^2}{\hbar^2}\!\! 
\int_0^{t} \!\!\! \!dt_1 \!\! \int_0^{t_1} \!\!\!\!\! \!dt_2 \!\!\int \!\! dr_1 dr_2 \, \overline{\phi}_1(\!r_1,\!k_1\!) \overline{\phi}_2(\!r_2,\!k_2\!) \left( W_j(t_1) W_l(t_2) \Psi_0 \right)\! (\!R,\! r_1,\! r_2)\label{calfjl}
\ea

\n
where $l,j=1,2$, $j\neq l$. Furthermore, using the specific factorized  form of the initial state of the system, we have

\ba\label{calf2}
&&\mathcal{F}_{lj}(R, \!k_1,\!k_2,\! t)  \!=\! \f{\lambda^2}{\hbar^2}\!\! 
\int_0^{t} \!\!\! \!dt_1 \!\! \int_0^{t_1} \!\!\!\!\! \!dt_2  \; e^{\f{i}{\hbar} t_1\left( E(k_j)+|E_0| \right)}  e^{\f{i}{\hbar} t_2\left( E(k_l) +|E_0| \right)}
   \left( e^{\f{i}{\hbar} t_1 K_0} \hat{V}_j (\cdot,k_j) e^{-\f{i}{\hbar} t_1 K_0}\right.\nonumber\\
&&
\left. e^{\f{i}{\hbar} t_2 K_0} \hat{V}_l (\cdot,k_l) e^{-\f{i}{\hbar} t_2 K_0}  \psi \right)\!(R) 
\ea

\n
where $E(k)=\f{\hbar^2 k^2}{2 M}$ and  $\hat{V}_j(\cdot,k_j)$, $j=1,2$, is the multiplication operator  by

\ba\label{vh}
&&\hat{V}_j (R,k_j) = \int \!\! dr \; \overline{\phi}_j (r,k_j) V(\gamma^{-1}(R-r) ) \zeta_j(r)
\ea

\n
The r.h.s. of (\ref{vh}) can be more conveniently written in terms of the Fourier transform $\widetilde{V}$ of the interaction potential as follows
\ba\label{vh1}
&&\hat{V}_j (R,k_j) = e^{-i k \cdot a_j} \, \gamma^{3/2}\! \int \!\! d\xi \; e^{i \gamma^{-1} (R- a_j) \cdot \xi} \;
g(\xi,  \gamma k_j) 
\ea
where $g(\xi,y)$ has been defined in (\ref{g}). 
Exploiting (\ref{vh1}) and the explicit expression of the free propagator we have

\ba\label{kvk}
&&\left( e^{\f{i}{\hbar} t_2 K_0} \hat{V}_l (\cdot,k_l) e^{-\f{i}{\hbar} t_2 K_0}  \psi \right)\!(R) \nonumber\\
&&= e^{-ik_l \cdot a_l} \, \gamma^{3/2}\! \!\int \!\! d\xi \, g(\xi,  \gamma k_l) \, e^{ i \f{\hbar t_2}{2 M \gamma^2} \xi^2   + i  \f{R}{\gamma} \cdot \xi    - i  \f{a_l}{\gamma} \cdot \xi} \, \psi \Big(\! R + \f{\hbar t_2}{M \gamma} \xi \! \Big) 
\ea

\n
and

\ba\label{kvk2}
&& \left( e^{\f{i}{\hbar} t_1 K_0} \hat{V}_j (\cdot,k_j) e^{-\f{i}{\hbar} t_1 K_0}  e^{\f{i}{\hbar} t_2 K_0} \hat{V}_l (\cdot,k_l) e^{-\f{i}{\hbar} t_2 K_0}  \psi \right)\!(R) \nonumber\\
&&=e^{-i k_l \cdot a_l - i k_j \cdot a_j} \, \gamma^3 \! \!\! \int \!\! d\xi d\eta \, g(\eta,\gamma k_j) g(\xi, \gamma k_l) \,\nonumber\\
&& \cdot \, e^{  i \left(  
\f{\hbar t_1}{2 M \gamma^2} \eta^2   +  \f{R}{\gamma} \cdot \eta   +  \f{\hbar t_2}{2 M \gamma^2} \xi^2   +  \f{R}{\gamma} \cdot  \xi  +  \f{\hbar t_1}{M \gamma^2} \eta \cdot \xi 
\right)} \, e^{ i \left( 
- \f{a_j}{\gamma} \cdot \eta -   \f{a_l}{\gamma} \cdot \xi 
\right)} 
\, \psi \Big(\! R + \f{\hbar t_1}{M\gamma} \eta +\f{\hbar t_2}{M \gamma} \xi \! \Big) 
\ea

\n
Finally we consider the time-dependent phase factor in (\ref{calf2}). We notice that

\ba\label{tph}
&&\f{t_1}{\hbar} (E(k_j)+|E_0|) + \f{t_2}{\hbar} (E(k_l)+|E_0|) 
= \f{\hbar}{m \gamma^2}   \left( t_1 w(\gamma k_j) + t_2  w(\gamma k_l) \right)
\ea

\n
Taking into account  (\ref{kvk2}), (\ref{tph}),  (\ref{sfe}), and rescaling the time variables according to $t_1 = t \alpha$, $t_2 = t \beta$, we can rewrite (\ref{calf2}) as follows 

\ba\label{calf3}
&&\mathcal{F}_{lj}(R,k_1,k_2,t)\nonumber\\
&&=  \f{\lambda^2 t^2}{\hbar^2} \f{\mathcal{N}}{\ve}  \gamma^{3/2}  e^{-i k_l \cdot a_l -i k_j \cdot a_j} \!\! \int_{S^2} \!\! d \hat{u} 
\! \!\int_0^1 \!\!\! d \alpha \!\! \int_0^{\alpha} \!\! \!d \beta \!\! \int \!\! d \xi d \eta \, g(\eta, \gamma k_j) g(\xi, \gamma k_l) f  \Big(\! \f{R}{\gamma} +\! \f{\hbar t}{M\gamma^2 } \!\left( \alpha \eta + \beta \xi \right) \!\! \Big)
\nonumber\\
&&\cdot \, e^{i  \left(   \f{\hbar t}{2 M \gamma^2} \alpha \eta^2   +  \f{R}{\gamma} \cdot \eta   +  \f{\hbar t}{2 M \gamma^2}\beta  \xi^2   +  \f{R}{\gamma} \cdot  \xi  + \f{\hbar t}{M \gamma^2} \alpha \eta \cdot \xi  \right)} \, 
e^{ i \left[ 
\f{1}{\ve} \hat{u} \cdot  \left( \f{R}{\gamma} 
 + \f{\hbar t}{M \gamma^2} \left(\alpha \eta + \beta \xi \right) \right)
- \f{a_j}{\gamma} \cdot \eta -   \f{a_l}{\gamma} \cdot \xi 
+ \f{\hbar t}{m \gamma^2} \left( \omega(\gamma k_j) \alpha + \omega(\gamma k_l) \beta \right) 
\right] }\nonumber\\
&&
\ea

\n
We observe that 
\ba\label{phl}
&&\f{\hbar t}{M \gamma^2} =  \ma = O(1)
\ea

\n
and this means that the first exponential in the integral in (\ref{calf3}) has a slowly oscillating phase for $\ve \ll 1$. On the other  hand
\ba\label{phr}
&&\f{\hbar t}{m \gamma^2} = \f{M}{m}  \ma =O(\ve^{-1})
\ea

\n
and therefore the last exponential in the integral in (\ref{calf3}) has a rapidly  oscillating phase for $\ve \ll 1$.  
Denoting  $R = \gamma x$, $k_j = \gamma^{-1} y_j$ and using the notation (\ref{frak}), 
we find

\ba\label{calf4}
&&\mathcal{F}_{lj}(\gamma x,\gamma^{-1}y_1, \gamma^{-1} y_2) \nonumber\\
&&=  \f{\lambda^2 t^2}{\hbar^2} \f{\mathcal{N}}{\ve}  \gamma^{3/2}  e^{-i k_l \cdot a_l -i k_j \cdot a_j} \!\! \int_{S^2} \!\! d \hat{u} 
\! \!\int_0^1 \!\!\! d \alpha \!\! \int_0^{\alpha} \!\! \!d \beta \!\! \int \!\! d \xi d \eta \,g(\eta,y_j) g(\xi,y_l) f(x+ \ma (\alpha \eta +\beta \xi)) \nonumber\\
&&e^{i \left[    x \cdot (\eta +  \xi  ) +
\f{ \mathfrak{a}}{2}  \,( \alpha \eta^2 + \beta \xi^2 + 2 \alpha \, \eta \cdot \xi ) \right]} \;  e^{ \f{i}{\ve} \left[    \hat{u} \cdot x + \hat{u} \cdot \ma(\alpha \eta + \beta \xi) 
  - \mathfrak{b}_j \hat{a}_j \cdot \eta - \mathfrak{b}_l \hat{a}_l \cdot \xi + \mathfrak{c}_j \alpha + \mathfrak{c}_l \beta  \right]   } \nonumber\\
&& \equiv  \f{\lambda^2 t^2}{\hbar^2} \f{\mathcal{N}}{\ve}  \gamma^{3/2} e^{-i k_l \cdot a_l -i k_j \cdot a_j} \mathcal{G}_{lj}^{\ve}  (x,y_1,y_2,t)
\ea

\n
where in the last line we have used (\ref{gjl}), (\ref{tejl}),  (\ref{gjl2}), (\ref{phi}). 
From (\ref{prio}), (\ref{calf}) and (\ref{calf4}) we obtain

\ba\label{p3}
&&\mathcal{P}(t)=  \gamma^{-3} \int \!\! dx dy_1  dy_2 \left| \mathcal{F}_{12} (\gamma x, \gamma^{-1} y_1 , \gamma^{-1} y_2 ) +  \mathcal{F}_{21} (\gamma x, \gamma^{-1} y_1 , \gamma^{-1} y_2 ) \right|^2 \nonumber\\
&&=  \f{\lambda^4 t^4}{\hbar^4} \f{ \mathcal{N}^2}{\ve^2} \int \!\! dx dy_1  dy_2 \left| \mathcal{G}_{12}^{\ve}(x,y_1,y_2) + \mathcal{G}_{21}^{\ve} (x,y_1,y_2) \right|^2
\ea

\n
and this concludes the proof of (\ref{p2}).


\vs
\vs
\vs
\vs
\vs

\vs
\vs
\vs

\end{document}